\documentclass[aps,pra,superscriptaddress]{revtex4}% Physical Review B
\usepackage{graphicx}% Include figure files
\usepackage{dcolumn}% Align table columns on decimal point
\usepackage{bm}% bold math
\usepackage{amsmath}
\usepackage{amssymb,amsbsy,cancel}
\usepackage{color,ulem}
\usepackage{here}
\usepackage{natbib}
\newcommand{\beq}{\begin{equation}}
\newcommand{\eeq}{\end{equation}}
\newcommand{\bea}{\begin{eqnarray}}
\newcommand{\eea}{\end{eqnarray}}
\newcommand{\bec}{\begin{center}}
\newcommand{\enc}{\end{center}}
\newcommand{\bfr}{\begin{flushright}}
\newcommand{\efr}{\end{flushright}}
\newcommand{\la}{\langle}
\newcommand{\ra}{\rangle}

\newcommand{\alp}{\alpha}

\newcommand{\om}{\omega}

\newcommand{\gam}{\gamma}

\newcommand{\Om}{\Omega}
\newcommand{\lam}{\lambda}

\newcommand{\hb}{\hat{b}}
\newcommand{\hs}{\hat{o}}
\newcommand{\hS}{\hat{S}}
\newcommand{\hH}{\hat{H}}
\newcommand{\hN}{\hat{N}}
\newcommand{\hsig}{\hat{\sigma}}
\newcommand{\tb}{\widetilde{b}}%\rm 
\newcommand{\tom}{\widetilde{\omega}}
\newcommand{\tOm}{\widetilde{\Omega}}

\newcommand{\cN}{{\cal N}}

\begin{document}
%%%%%%%%%%%%%%%%%%%%%%%%%%%%%%%%%%%%%%%%%%%%%%%%%%%%%%%%%%%%%%%%%%%%%%%%%%%%%%%
%%%%%%%%%%%%%%%%%%%%%%%%%%%%%%%%%%%%%%%%%%%%%%%%%%%%%%%%%%%%%%%%%%%%%%%%%%%%%%%
\title{
% Theory of Josephson quantum filter
%
% Protection of a qubit with a highly dissipative qubit:\\
%
Protection of a qubit with a subradiance effect: Josephson quantum filter
}
%%%%%%%%%%%%%%%%%%%%%%%%%%%%%%%%%%%%%%%%%%%%%%%%%%%%%%%%%%%%%%%%%%%%%%%%%%%%%%%
%%%%%%%%%%%%%%%%%%%%%%%%%%%%%%%%%%%%%%%%%%%%%%%%%%%%%%%%%%%%%%%%%%%%%%%%%%%%%%%
\author{Kazuki Koshino}
\email{kazuki.koshino@osamember.org}
\affiliation{College of Liberal Arts and Sciences, Tokyo Medical and Dental
University, Ichikawa, Chiba 272-0827, Japan}
\author{Shingo Kono}
\affiliation{RIKEN Center for Emergent Matter Science (CEMS), 2-1 Hirosawa, Wako, 
Saitama 351-0198, Japan}
\author{Yasunobu Nakamura}
\affiliation{RIKEN Center for Emergent Matter Science (CEMS), 2-1 Hirosawa, Wako, 
Saitama 351-0198, Japan}
\affiliation{Research Center for Advanced Science and Technology (RCAST), 
The University of Tokyo, Meguro-ku, Tokyo 153-8904, Japan}
%%%%%%%%%%%%%%%%%%%%%%%%%%%%%%%%%%%%%%%%%%%%%%%%%%%%%%%%%%%%%%%%%%%%%%%%%%%%%%%
%%%%%%%%%%%%%%%%%%%%%%%%%%%%%%%%%%%%%%%%%%%%%%%%%%%%%%%%%%%%%%%%%%%%%%%%%%%%%%%
\date{\today}
%%%%%%%%%%%%%%%%%%%%%%%%%%%%%%%%%%%%%%%%%%%%%%%%%%%%%%%%%%%%%%%%%%%%%%%%%%%%%%%
%%%%%%%%%%%%%%%%%%%%%%%%%%%%%%%%%%%%%%%%%%%%%%%%%%%%%%%%%%%%%%%%%%%%%%%%%%%%%%%
\begin{abstract}
The coupling between a superconducting qubit and a control line 
inevitably results in radiative decay of the qubit into the line. 
We propose a Josephson quantum filter (JQF), 
which protects the data qubit (DQ) from radiative decay 
through the control line
without reducing the gate speed on DQ. 
JQF consists of a qubit strongly coupled to the control line to DQ, 
and its working principle is a subradiance effect 
characteristic to waveguide quantum electrodynamics setups. 
JQF is a passive circuit element and is therefore
suitable for integration in a scalable superconducting qubit system. 
\end{abstract}
%%%%%%%%%%%%%%%%%%%%%%%%%%%%%%%%%%%%%%%%%%%%%%%%%%%%%%%%%%%%%%%%%%%%%%%%%%%%%%%
% \pacs{
% 42.50.Ar, % Photon statistics and coherence theory 
% 42.50.Pq, % Cavity quantum electrodynamics; micromasers 
% 42.65.Sf % optical spatio-temporal dynamics 
% }
\maketitle

%%%%%%%%%%%%%%%%%%%%%%%%%%%%%%%%%%%%%%%%%%%%%%%%%%%%%%%%%%%%%%%%%%%%%%%%%%%%%%%
\section{Introduction}\label{sec:intro}
%%%%%%%%%%%%%%%%%%%%%%%%%%%%%%%%%%%%%%%%%%%%%%%%%%%%%%%%%%%%%%%%%%%%%%%%%%%%%%%

Superconducting qubit system has high designability 
and in-situ tunability of the system parameters
and is therefore suitable for realization of scalable quantum computation. 
Supported by recent progress in integrating superconducting qubits, 
quantum computers including several tens of qubits are currently
available~\cite{Barends2014,kelly2015,reagor2018,song2019,wei2019,ye2019}. 
In order to perform quantum computational tasks involving many qubits, 
we need to apply fast gate operations on the qubits, % and with high fidelities, 
while keeping the long coherence times of the qubits. 
These requirements are, however, conflicting usually. 
For gate operations on a superconducting qubit, 
we couple the qubit to a control line 
through which microwave gate pulses are applied. 
% Similarly, for the dispersive readout of the qubit state, 
% we couple the qubit is to a resonator and then to a control line. 
A strong coupling between the qubit and the control line 
is advantageous for the gate speed, 
but is disadvantageous for the qubit lifetime
due to radiative decay through the line. 
% Such control lines have continuous mechanical degrees of freedom 
% and cause radiative decay of the qubit. 
When the radiative decay rate of the qubit is $\gam$, 
the Rabi frequency induced by a drive field applied through the line
scales as $\gam^{1/2}$ for a fixed drive power. 
Therefore, the gate speed and the qubit lifetime
are proportional to $\gam^{1/2}$ and $\gam^{-1}$, respectively. 
A usual strategy to simultaneously realize 
a long qubit lifetime and fast gate operations
is to make $\gam$ %the coupling between DQ and control line 
small for enhancing the qubit lifetime
and to apply short and intense pulses for reducing the gate time. 
For example, we can halven the gate time
by doubling the amplitude of the control pulse.
This results in, however, doubling of the average photon number per pulse
and resultantly more heating of the % \sout{qubit} system.
surrounding components such as attenuators and filters passing the pulse.
Intense control pulses may also induce unwanted 
% excitation to the second excited state of the qubit.
crosstalk with neighboring qubits and resonators.

In the dispersive readout of qubits~\cite{blais2004,vijay2011}, 
a similar trade-off exists between 
the measurement speed and the qubit lifetime. 
In this scheme, 
we couple the qubit dispersively to a resonator and then to a readout line, 
through which readout microwave pulses are applied. 
A strong coupling between the resonator and the readout 
line enables fast measurements, but shortens the qubit lifetime
due to resonator-mediated radiative decay (Purcell effect)~\cite{bloembergen1948}. 
Here, the readout frequency is essentially that of the resonator, 
whereas the frequency of an emitted photon is that of the qubit. 
% This attached waveguide causes spontaneous emission of the qubit
% via the Purcell effect and shortens the qubit lifetime.
Using this frequency difference, 
we can resolve the trade-off by incorporating a frequency filter (Purcell filter)
in the readout line~\cite{reed2010,jeffrey2014,Sete2015,walter2017}. 
However, such frequency filtering is inapplicable 
for suppression of decay into the control line, 
since control pulses have the same frequency as the leaking photon.

%==============================================================================
\begin{figure} % [b]
\begin{center}
\includegraphics[width=70mm]{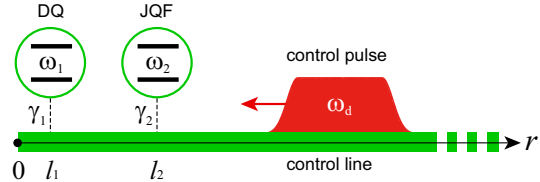}
\end{center}
\caption{Schematic of the setup. 
The data qubit (DQ) and the Josephson quantum filter (JQF) 
are directly coupled to a semi-infinite waveguide, 
through which control pulses for DQ is applied. 
Unless otherwise specified, 
the following parameter values are assumed:
$\om_1/2\pi=\om_2/2\pi=5$~GHz
(resonance wavelength $\lam_q=20$~mm, 
for the microwave velocity $v=10^8$~m/s), 
$l_1=0$~mm,
$l_2=\lam_q/2=10$~mm,
$\gam_1/2\pi=2$~kHz, and 
$\gam_2/2\pi=100$~MHz.
}
\label{fig:config}
\end{figure}
%==============================================================================

In this study, 
we propose a filter % Josephson quantum filter (JQF)
that prohibits radiative decay of the qubit into the contorl line 
while not affecting the control speed. 
Figure~\ref{fig:config} is the schematic of the considered setup:
a data qubit (DQ) to be controlled is coupled 
to an end of a semi-infinite control line,   
and a filter qubit, % (Josephson quantum filter, JQF)
which is referred to as the Josephson quantum filter (JQF), 
is placed at a distance of the order of the resonance wavelength of DQ. 
% We attach a semi-infinite control line to a data qubit (DQ). 
% In front of DQ, we strongly couple another qubit of the same frequency,
In such waveguide-QED setups,
it is known that a qubit functions as a nonlinear mirror,  
which completely reflects a weak field due to destructive interference
and transmits a stronger field due to absorption 
saturation~\cite{shen2005a,shen2005b,astafiev2010}. 
In radiative decay of DQ, where only a single photon concerns, 
JQF works as a mirror  
% forms an effective cavity with the termination point of the control line
that prohibits radiative decay of DQ. 
On the other hand, 
when we perform gate operations on DQ by applying strong control pulses,
JQF transmits the pulses and does not reduce the gate speed. 
Since JQF is a passive circuit element which is free from active control on it, 
JQF is ready to be incorporated in complicated circuits involving many qubits. 
Together with the established schemes such as 
Purcell filters~\cite{reed2010,jeffrey2014,Sete2015,walter2017}
and tunable qubit-waveguide 
couplers~\cite{sete2013catch,bockstiegel2016tunable,forn2017demand}, 
JQF would be highly useful for constructing a network of long-lived qubits 
yet allowing fast control and measurements.

The rest of this paper is organized as follows.
In Sec.~\ref{sec:form}, we theoretically describe the considered setup.
We present the Hamiltonian of the overall system % which is 
composed of DQ, JQF and a semi-infinite control line, 
and derive their Heisenberg equations.
In Sec.~\ref{sec:to}, 
assuming the absence of JQF, 
we review the trade-off between the gate speed and the lifetime of DQ.
In Sec.~\ref{sec:rd}, 
we analyze radiative decay of initially excited DQ. 
We derive an analytic formula of the radiative decay rate of DQ
and clarify the condition that JQF protects DQ from radiative decay.
In Sec.~\ref{sec:mr}, 
we numerically examine the microwave response of DQ and JQF
to confirm the following: 
JQF does not affect the dynamics of DQ induced by the control field
and therefore does not reduce the gate speed.
At the same time, 
JQF prohibits radiative decay of DQ while the control field is off. 
Section~\ref{sec:sum} is devoted to the summary.

%%%%%%%%%%%%%%%%%%%%%%%%%%%%%%%%%%%%%%%%%%%%%%%%%%%%%%%%%%%%%%%%%%%%%%%%%%%%%%%
\section{Formulation}\label{sec:form}
%%%%%%%%%%%%%%%%%%%%%%%%%%%%%%%%%%%%%%%%%%%%%%%%%%%%%%%%%%%%%%%%%%%%%%%%%%%%%%%
%%%%%%%%%%%%%%%%%%%%%%%%%%%%%%%%%%%%%%%%%%%%%%%%%%%%%%%%%%%%%%%%%%%%%%%%%%%%%%%
\subsection{Hamiltonian}%\label{sec:Ham}
%%%%%%%%%%%%%%%%%%%%%%%%%%%%%%%%%%%%%%%%%%%%%%%%%%%%%%%%%%%%%%%%%%%%%%%%%%%%%%%
The physical setup considered in this study is 
schematically illustrated in Fig.~\ref{fig:config}.
In order to apply control microwave pulses to DQ (qubit~1), 
we attach a semi-infinite waveguide to DQ.
In front of DQ, we attach JQF (qubit~2), 
which is also a qubit with the same frequency as DQ.
The semi-infinite waveguide extends in the $r>0$ region.
Its eigenmodes are standing waves, 
which are continuously labeled by a wavenumber $k(>0)$. 
Assuming the open boundary condition at the termination point ($r=0$), 
the mode function $f_k(r)$ is given by
\bea
f_k(r) &=& \sqrt{\frac{2}{\pi}}\cos kr.
\label{eq:cos}
\eea
We denote the annihilation operator for this mode by $\hb_k$. 
The mode functions are normalized as 
$\int_0^{\infty} dr f^*_{k'}(r)f_k(r)=\delta(k-k')$.

Both qubits can be regarded as two-level systems.
We denote the lowering operator of qubit~$m$ ($m=1,2$) by $\hsig_m$,
the transition frequency by $\om_m$, 
the coupling position to the waveguide by $l_m$ 
and the coupling strength by $\gam_m$. 
Note that $\gam_m$ represents 
the radiative decay rate of qubit~$m$, % to the waveguide, 
assuming that the qubit is coupled to an {\it infinite} waveguide.

Setting $\hbar=v=1$, 
where $v$ is the microwave velocity in the waveguide,
the Hamiltonian of the overall system
composed of DQ, JQF, and the semi-infinite waveguide 
is given by
% \bea
% \cH &=& \om_1 \sig_1^{\dag}\sig_1 + \om_2 \sig_2^{\dag}\sig_2 
% + \int dk \ \left[ kb_k^{\dag}b_k + 
% \right.
% \nonumber 
% \\
% & & 
% \left.
% g_{1k}(\sig_1^{\dag}b_k + b_k^{\dag}\sig_1)
% + g_{2k}(\sig_2^{\dag}b_k + b_k^{\dag}\sig_2) \right],
% \eea
\bea
\hat{H} &=& 
\sum_m \om_m \hsig_m^{\dag}\hsig_m
+ \int_0^{\infty} dk \ \left[ k\hb_k^{\dag}\hb_k 
+ \textstyle{\sum}_m
g_{mk}(\hsig_m^{\dag}\hb_k + \hb_k^{\dag}\hsig_m)\right],
\label{eq:Ham}
\eea
where the coupling constant $g_{mk}$ is given by
\bea
g_{mk} &=& \sqrt{\frac{\gam_m}{2}} f_k(l_m)
= \sqrt{\frac{\gam_m}{\pi}}\cos(kl_m).
\eea

%%%%%%%%%%%%%%%%%%%%%%%%%%%%%%%%%%%%%%%%%%%%%%%%%%%%%%%%%%%%%%%%%%%%%%%%%%%%%%%
% \subsection{system parameters}%\label{sec:Ham}
%%%%%%%%%%%%%%%%%%%%%%%%%%%%%%%%%%%%%%%%%%%%%%%%%%%%%%%%%%%%%%%%%%%%%%%%%%%%%%%
In this study, we aim to protect DQ 
from radiative decay using JQF. 
For this purpose, the system parameters are chosen as follows.
(i)~The two qubits are nearly resonant,
i.e., $\om_1 \approx \om_2 \approx \om_q$.
(ii)~The positions $l_1$ and $l_2$ of the qubits
are of the order of their resonance wavelength,
$\lam_q = 2\pi v/\om_q$. 
DQ is closer to the end of the waveguide than JQF, i.e., $0 \leq l_1 \leq l_2$. 
(iii)~The coupling between JQF and the waveguide
is much stronger than that of DQ,
i.e., $\gam_1 \ll \gam_2$. 
Unless otherwise specified, 
we employ the parameter values listed in the caption of Fig.~\ref{fig:config}.

%==============================================================================
% \begin{table}[b]
% \caption{\label{tab:table1}%
% Typical parameter values.
% }
% \begin{ruledtabular}
% \begin{tabular}{cc}
% $\om_1/2\pi$ & 5~GHz \\
% $\om_1/2\pi$ & 5~GHz \\
% $\om_1/2\pi$ & 5~GHz \\
% $\om_1/2\pi$ & 5~GHz 
% \end{tabular}
% \end{ruledtabular}
% \end{table}
%==============================================================================

For a semi-infinite waveguide, 
% the eigen waveguide modes are the standing waves, 
the wavenumber $k$ of the waveguide mode is restricted to be positive.
However, we formally extend 
the lower limit of $k$ to $-\infty$
\footnote{
This introduces the modes with negative energies. 
However, such modes are highly detuned from the qubits
and have little effect on the dynamics. 
},
and introduce the real-space representation $\tb_r$ of 
the field operator by
\bea
\tb_r &=& \frac{1}{\sqrt{2\pi}} 
\int_{-\infty}^{\infty} dk \ e^{ikr}\hb_k.
\label{eq:br}
\eea
The space variable $r$ runs over $-\infty<r<\infty$: 
the negative (positive) region represents the incoming (outgoing) field.
The introduction of the real-space representation
has been validated by a rigorous 
^^ ^^ modes of the universe'' approach~\cite{gea2013space}.

%%%%%%%%%%%%%%%%%%%%%%%%%%%%%%%%%%%%%%%%%%%%%%%%%%%%%%%%%%%%%%%%%%%%%%%%%%%%%%%
\subsection{Heisenberg equations}%\label{sec:Ham}
%%%%%%%%%%%%%%%%%%%%%%%%%%%%%%%%%%%%%%%%%%%%%%%%%%%%%%%%%%%%%%%%%%%%%%%%%%%%%%%
The Heisenberg equation for $\hb_k$ is given by
$\frac{d}{dt}\hb_k=-ik\hb_k-i\sum_m g_{mk}\hsig_m$. 
This is formally solved as
\bea
\hb_k(t) &=& \hb_k(0)e^{-ikt}-i\sum_m g_{mk} \int_0^t dt' \hsig_m(t')e^{ik(t'-t)}.
\eea
Switching to the real-space representation with Eq.~(\ref{eq:br}), 
% From the Heisenberg equations for the field operator $\hb_k$, 
% which is obtained from Eq.~(\ref{eq:Ham}),
we have % obtain the following equality: 
\bea
\tb_r(t) &=& \tb_{r-t}(0)-i\sum_m\sqrt{\frac{\gam_m}{2}}
\left[
\Theta_{r \in (-l_m,t-l_m)}\hsig_m(t-r-l_m) + 
\Theta_{r \in ( l_m,t+l_m)}\hsig_m(t-r+l_m)
\right],
\label{eq:tbrt}
\eea
where $\Theta$ is a product of the step functions,
$\Theta_{r \in (a,b)}=\theta(r-a)\theta(b-r)$.
% where $\theta$ is the step function.
This equation plays the role of the input-output relation 
in quantum optics~\cite{gardiner2004,walls2007}.

The Heisenberg equation for a system operator $\hs$ 
(composed of $\hsig_1$, $\hsig_2$ and their conjugates)
is written, from Eqs.~(\ref{eq:Ham}) and (\ref{eq:br}), as
\bea
\frac{d}{dt}\hs 
&=&
i[\hH_s, \hs] + 
i\sum_m \sqrt{\frac{\gam_m}{2}}\left(
[\hsig_m^{\dag},\hs] \left\{ \tb_{l_m}(t) + \tb_{-l_m}(t) \right\}
+
\left\{ \tb_{l_m}^{\dag}(t) + \tb_{-l_m}^{\dag}(t) \right\} [\hsig_m,\hs]
\right),
\label{eq:dsdt5}
\eea
where $\hH_s=\sum_m \om_m \hsig_m^{\dag} \hsig_m$. 
From Eq.~(\ref{eq:tbrt}), we obtain the following equality:
%$\tb_{l_m}(t) + \tb_{-l_m}(t)$ is rewritten as
\bea
\tb_{l_m}(t) + \tb_{-l_m}(t) &=& \tb_{l_m-t}(0) + \tb_{-l_m-t}(0)
-i\sum_n \sqrt{\frac{\gam_n}{2}} \left[ \hsig_n(t-l_m-l_n)+\hsig_n(t-|l_m-l_n|) \right].
\label{eq:tblmt}
\eea
From Eqs.~(\ref{eq:dsdt5}) and (\ref{eq:tblmt}), we have
\bea
\frac{d}{dt}\hs 
&=&
i[\hH_s, \hs] + 
i\sum_m\left\{
[\hsig_m^{\dag}, \hs]\hN_m(t) + \hN^{\dag}_m(t)[\hsig_m, \hs]
\right\}
\nonumber
\\
&+& \sum_{m,n}\frac{\sqrt{\gam_m\gam_n}}{2}[\hsig_m^{\dag},\hs]
\left\{
\hsig_n(t-l_m-l_n)+\hsig_n(t-|l_m-l_n|)
\right\}
\nonumber
\\
&-& \sum_{m,n}\frac{\sqrt{\gam_m\gam_n}}{2}
\left\{
\hsig_n^{\dag}(t-l_m-l_n)+\hsig_n^{\dag}(t-|l_m-l_n|)
\right\}
[\hsig_m,\hs].
\label{eq:dsjdt}
\eea
% \sout{qubit~$\overline{j}$ is the opposite of qubit~$j$ ($\overline{j}=3-j$),}
Note that the explicit time dependence is omitted for operators at time $t$,
and the operators with negative time variable should be replaced with zero.
The noise operator $\hat{N}_m(t)$ for qubit $m$ is defined by 
\bea
\hat{N}_m(t) &=& 
\sqrt{\frac{\gam_m}{2}}\left[
\tb_{l_m-t}(0)+\tb_{-l_m-t}(0)
\right].
\eea

%%%%%%%%%%%%%%%%%%%%%%%%%%%%%%%%%%%%%%%%%%%%%%%%%%%%%%%%%%%%%%%%%%%%%%%%%%%%%%%
\subsection{Free-evolution approximation}%\label{sec:Ham}
%%%%%%%%%%%%%%%%%%%%%%%%%%%%%%%%%%%%%%%%%%%%%%%%%%%%%%%%%%%%%%%%%%%%%%%%%%%%%%%
The Heisenberg equation (\ref{eq:dsjdt}), % \sout{for $\hsig_j(t)$}, 
which is rigorously driven from the Hamiltonian of Eq.~(\ref{eq:Ham}),
contains qubit operators with retarded times. 
However, considering that 
the qubit decay during such retardation is negligibly small
($\gam_j l_j/v \ll 1$), 
we can employ the free-evolution approximation~\cite{johansson2008,koshino2012},
$\hsig_j(t-\Delta t) \approx e^{i\om_q\Delta t}\hsig_j(t)$. 
The validity of this approximation is confirmed in Appendix~\ref{sec:app}.
Then, Eq.~(\ref{eq:dsjdt}) is recast into the following simpler form:
\bea
\frac{d}{dt}\hs 
&=&
i[\hH_s, \hs] + 
i\sum_m\left\{
[\hsig_m^{\dag}, \hs]\hN_m(t) + \hN^{\dag}_m(t)[\hsig_m, \hs]
\right\}
% \nonumber \\ &+& 
+ \sum_{m,n} \left(
\xi_{mn}[\hsig_m^{\dag},\hs]\hsig_n - \xi^*_{mn}\hsig_n^{\dag}[\hsig_m,\hs]
\right),
\label{eq:dsdt}
\\
\xi_{mn} &=& \frac{\sqrt{\gam_m\gam_n}}{2}
\left( e^{i\om_q(l_m+l_n)}+e^{i\om_q|l_m-l_n|} \right).
\label{eq:ximn}
\eea
% &+& \sum_j \left(
% \xi_{jj} [\hsig_j^{\dag}, \hs]\hsig_j -
% \xi_{jj}^* \hsig_j^{\dag}[\hsig_j, \hs] +
% \xi_{j\oj} [\hsig_j^{\dag}, \hs]\hsig_{\oj} -
% \xi_{j\oj}^* \hsig_{\oj}^{\dag}[\hsig_j, \hs] \right),
% \\
% \xi_{jj} &=& \gam_j e^{i\om_q l_j}\cos(\om_q l_j),
% \\
% \xi_{j\overline{j}} &=& \sqrt{\gam_1\gam_2} e^{i\om_q l_2}\cos(\om_q l_1).
Although the direct interaction between DQ and JQF is absent 
in the Hamiltonian of Eq.~(\ref{eq:Ham}), 
virtual waveguide photons mediate 
an effective interaction between them, % \cite{chang2012,lalumiere2013,van2013}.
which appears as 
$\xi_{12}=\xi_{21}=\sqrt{\gam_1\gam_2}\cos(\om_ql_1)e^{i\om_ql_2}$.
Note that the roles of DQ and JQF are asymmetric in this interaction
due to the semi-infiniteness of the control line. 

Since the detuning of the control pulse 
from the qubit resonance, $\om_d-\om_q$, 
is small enough to satisfy $(\om_d-\om_q)l_j/v \ll 1$
in the considered setup,
we can make the following approximation, 
$\tb_{\Delta r -t}(0) \approx e^{i\om_q \Delta r}
\tb_{-t}(0)$.
Then we have
\bea
\hat{N}_j(t) 
&=& 
\sqrt{2\gam_j}\cos(\om_ql_j)\tb_{-t}(0).
\label{eq:Njt}
\eea
For future reference, setting $\hs = \hsig_1$, we have
\bea
\frac{d}{dt}\hsig_1 &=& (-i\om_q-\xi_{11})\hsig_1 
-\xi_{12}(1-2\hsig_1^{\dag}\hsig_1)\hsig_2
+i(1-2\hsig_1^{\dag}\hsig_1)\hN_1(t).
\label{eq:dsig1dt}
\eea

%%%%%%%%%%%%%%%%%%%%%%%%%%%%%%%%%%%%%%%%%%%%%%%%%%%%%%%%%%%%%%%%%%%%%%%%%%%%%%%
\subsection{Effective interaction and cooperative dissipator}%\label{sec:Ham}
%%%%%%%%%%%%%%%%%%%%%%%%%%%%%%%%%%%%%%%%%%%%%%%%%%%%%%%%%%%%%%%%%%%%%%%%%%%%%%%
In the preceding theoretical works dealing with the qubit system
coupled to an infinite waveguide~\cite{chang2012,lalumiere2013,van2013,konyk2017}, 
the photon-mediated qubit-qubit interaction
is treated by the effective dipole exchange interaction $\hH_e$
and the cooperative dissipator $\hS$. 
Here, % in order to observe the connection to such formulation, 
we rewrite Eqs.~(\ref{eq:dsdt}) and (\ref{eq:ximn}) in this form for reference. 
Dividing $\xi_{mn}$ into the real and imaginary parts, we have
\bea
\frac{d}{dt}\hs 
&=&
i[\hH_s+\hH_e, \hs] 
+ \frac{ [\hS^{\dag}, \hs] \hS + \hS^{\dag} [\hs, \hS] }{2}
+ i\left\{ \hb_{\mathrm{in}}^{\dag}(t)[\hS,\hs]+[\hS^{\dag},\hs]\hb_{\mathrm{in}}(t) \right\},
\\
\hH_e &=& \sum_{m,n}J_{mn}\hsig_m^{\dag}\hsig_n,
\\
\hS &=& \sum_m \sqrt{2\gam_m}\cos(\om_q l_m)\hsig_m,
\\
J_{mn} &=& \sqrt{\gam_m\gam_n}\cos[\om_0\min(l_m,l_n)]\sin[\om_0\max(l_m,l_n)].
\eea
where $\hb_{\mathrm{in}}(t)=\tb_{-t}(0)$. 
Three comments are in order regarding the above equations.
(i)~The cooperative dissipator $\hS$ and 
the effective interaction $\hH_e$ result from 
the real and imaginary parts of $\xi_{mn}$, respectively.
(ii)~In the case of an infinite waveguide, 
there appears two kinds of cooperative dissipators
corresponding to the positively and negatively propagating modes. 
Here, the cooperative dissipator $\hS$ is unique, 
and the cosine function appearing in $\hS$
results from the standing-wave mode function [Eq.~(\ref{eq:cos})]. 
(iii)~In the case of an infinite waveguide, 
the coefficient $J_{mn}$ of the effective interaction
depends only on the mutual distance between the qubits, $|l_m-l_n|$. 
Here, $J_{mn}$ has a more complicated form
due to the semi-infiniteness of the waveguide.
% The semi-infiniteness of the present setup is reflected in 
% (i)~the cosine function appearing in $\hS$, 
% which originates in 
% the standing-wave mode function [Eq.~(\ref{eq:cos})],
% and (ii)~the coefficient $J_{mn}$ of the effective interaction, 
% which does not depend only on the mutual distance, $|l_m-l_n|$.  

%%%%%%%%%%%%%%%%%%%%%%%%%%%%%%%%%%%%%%%%%%%%%%%%%%%%%%%%%%%%%%%%%%%%%%%%%%%%%%%
\section{Trade-off between qubit lifetime and gate speed}
\label{sec:to}
%%%%%%%%%%%%%%%%%%%%%%%%%%%%%%%%%%%%%%%%%%%%%%%%%%%%%%%%%%%%%%%%%%%%%%%%%%%%%%%
In this section, we 
% observe the case in which the filter qubit is absent and 
observe the trade-off between the lifetime of DQ and the gate speed, 
assuming the absence of JQF ($\gam_2=0$).
The Heisenberg equation (\ref{eq:dsig1dt}) for $\hsig_1$ is then rewritten as
\bea
\frac{d}{dt}\hsig_1 &=& (-i\overline{\om}_q-\eta^2/2)\hsig_1 
+i\eta(1-2\hsig_1^{\dag}\hsig_1)\tb_{-t}(0),
\label{eq:dsig1dt2}
\eea
where $\overline{\om}_q = \om_q + (\gam_1/2)\sin(2\om_q l_1)$
is the renormalized qubit frequency including the Lamb shift.
A real constant $\eta=\sqrt{2\gam_1}\cos(\om_q l_1)$
has two roles that originate in the fluctuation-dissipation theorem:
the radiative decay rate of the qubit and 
the coupling between the qubit and the applied field.

In the absence of the input field,
the qubit excitation probability decays as
$\frac{d}{dt} \la\hsig_1^{\dag}\hsig_1\ra = 
-\eta^2 \la\hsig_1^{\dag}\hsig_1\ra$.
Therefore, the qubit radiative lifetime $T_r$ is given by
\bea
T_r &=& \frac{1}{\eta^2}.
\label{eq:Tr}
\eea
Note that, when coupled to a semi-infinite waveguide, 
the radiative lifetime depends on the qubit position $l_1$
due to the broken translation symmetry.
On the other hand, 
when we apply a resonant control field 
$E_\mathrm{in}(t)=E_d e^{-i\overline{\om}_q t}$
through the waveguide, 
the qubit excitation probability evolves as
% a resonant control field 
% ${\cal E}(t)=E_\mathrm{in}e^{-i\overline{\om}_q t}$
% applied through the waveguide
% induces the Rabi oscillation of the qubit.
$\frac{d^2}{dt^2} \la\hsig_1^{\dag}\hsig_1\ra = 
2\eta^2|E_d|^2 (1-2\la\hsig_1^{\dag}\hsig_1\ra)$,
and exhibits the Rabi oscillation with the Rabi frequency 
of $\Omega_R=2\eta|E_d|$. 
Therefore, the gate speed $T_g^{-1}$ for %the NOT gate ($\pi$-pulse) 
a $\pi$-pulse is given by
\bea
% \frac{1}{T_g} 
T_g^{-1} &=& \frac{2\eta|E_d|}{\pi}.
\label{eq:Tg}
\eea

Thus, there exists a trade-off between the qubit radiative lifetime 
and the gate speed. 
The overall lifetime $T_1$ of the qubit 
is always shorter than the radiative lifetime $T_r$
due to other relaxation channels. 
Therefore, from Eqs.~(\ref{eq:Tr}) and (\ref{eq:Tg}), 
we have the following inequality,
\bea
T_1 \left(\frac{1}{T_g}\right)^2 \leq \frac{4|E_d|^2}{\pi^2}.
\label{eq:tradeoff}
\eea
$|E_d|^2$ in the right-hand side represents 
the photon rate of the applied field. 
This is limited, besides the practical reasons,
by the finite anharmonicity of the qubit,
which is particularly small for the transmon-type qubits.

%%%%%%%%%%%%%%%%%%%%%%%%%%%%%%%%%%%%%%%%%%%%%%%%%%%%%%%%%%%%%%%%%%%%%%%%%%%%%%%
\section{Radiative decay}\label{sec:rd}
%%%%%%%%%%%%%%%%%%%%%%%%%%%%%%%%%%%%%%%%%%%%%%%%%%%%%%%%%%%%%%%%%%%%%%%%%%%%%%%
In this section, we analytically investigate 
the radiative decay of DQ in the presence of JQF
and clarify the optimal condition for JQF.
As the initial state,
we consider a state in which only DQ is excited.
The state vector is written as
\bea
|\psi(0)\ra &=& \hsig_1^{\dag}|v\ra,
\label{eq:inivec}
\eea
where $|v \ra$ represents the vacuum state of the whole setup. 
% in which both qubits are in the ground state and 
% no propagating photons are present in the waveguide.
% in which no excitations are present in the whole setup.
Since the Hamiltonian of Eq.~(\ref{eq:Ham}) conserves the total excitation number, 
the state vector at time $t$,
$|\psi(t)\ra=e^{-i\hH t}|\psi(0)\ra$, 
is written as follows:
\bea
|\psi(t)\ra &=& 
\sum_{j=1,2}
\alp_j(t)\hsig_j^{\dag}|v\ra 
% \alp_1(t)\hsig_1^{\dag}|v\ra 
% + \alp_2(t)\hsig_2^{\dag}|v\ra 
+ \int dr \ f(r,t)\tb_r^{\dag}|v \ra, 
\label{eq:tvec}
\eea
where the coefficients $\alp_j(t)$
and the wavepacket of the emitted photon $f(r,t)$
satisfy the normalization condition of 
$\sum_j|\alp_j(t)|^2 + \int_0^t dr|f(r,t)|^2=1$.
Using the fact that $\hH|v\ra=0$, we have
$\alp_j(t)=\la v|\hsig_j|\psi(t)\ra
=\la v|\hsig_j(t)\hsig_1^{\dag}(0)|v\ra$. 
% Similarly, $\beta(t)=\la v|\hsig_2(t)\hsig_1^{\dag}(0)|v\ra$. 
From Eq.~(\ref{eq:dsig1dt}) and its counterpart for $\hsig_2(t)$, 
the equations of motion for $\alp_j(t)$ are given by
\bea
\frac{d\alp_1}{dt} 
&=& 
-(i\om_q+\xi_{11})\alp_1 -\xi_{12}\alp_2,
\label{eq:da1dt}
\\
\frac{d\alp_2}{dt} 
&=& 
-\xi_{21}\alp_1 -(i\om_q+\xi_{22})\alp_2. 
\label{eq:da2dt}
% \\
% g(r,t) &=& 
% -i\sqrt{2\gam_1}\cos(\om_1 l_1)\alp(t-r) -i\sqrt{2\gam_2}\cos(\om_2 l_2)\beta(t-r)
\eea
with the initial conditions of $\alp_1(0)=1$ and $\alp_2(0)=0$. 
In deriving the above equations, 
we used $\hN_j(t)|\psi(0)\ra=0$ and 
$\hsig_1(t)\hsig_2(t)\hsig_j^{\dag}(0)|v\ra=0$. 
For reference, the wavepacket of the emitted photon is given by
% \bea
% g(r,t) &=& 
% \begin{cases}
% -i\sqrt{2\gam_1}\cos(\om_q l_1)\alp_1(t-r) 
% -i\sqrt{2\gam_2}\cos(\om_q l_2)\alp_2(t-r)
% & (0 \leq r \leq t)
% \\
% 0 & ({\rm otherwise})
% \end{cases}.
% \eea
$f(r,t)=-i\sqrt{2\gam_1}\cos(\om_q l_1)\alp_1(t-r)
-i\sqrt{2\gam_2}\cos(\om_q l_2)\alp_2(t-r)$ for $0 \leq r \leq t$, 
and $f(r,t)=0$ otherwise.

%%%%%%%%%%%%%%%%%%%%%%%%%%%%%%%%%%%%%%%%%%%%%%%%%%%%%%%%%%%%%%%%%%%%%%%%%%%%%%%
\subsection{Decay rate of DQ}%\label{sec:Ham}
%%%%%%%%%%%%%%%%%%%%%%%%%%%%%%%%%%%%%%%%%%%%%%%%%%%%%%%%%%%%%%%%%%%%%%%%%%%%%%%
In the absence of the photon-mediated interaction between qubits,
the complex frequency of qubit~$j$ is given by
$\tom_j=\om_q-i\xi_{jj}$,
and the radiative decay rate is given by
$-2\mathrm{Im}(\tom_j)=2\gam_j \cos^2(\om_q l_j)$.
Therefore, except for the special case of 
$l_2 \approx \frac{1+2n}{4}\lam_q$ ($n=0, 1, \cdots$),
where $\lam_q=2\pi v/\om_q$ is the resonance wavelength of the qubits, 
JQF decays much faster than DQ.
Then, by switching to the frame rotating at $\om_q$ and 
applying the adiabatic approximation, Eq.~(\ref{eq:da2dt}) reduces to
$\alp_2(t) \approx -(\xi_{21}/\xi_{22})\alp_1(t)$. 
Substituting this into Eq.~(\ref{eq:da1dt}), 
we confirm that the complex frequency $\tom_1'$ 
of DQ reduces to a real quantity:
\bea
\tom_1' 
% &=& \om_q - i(\xi_{11}-\xi_{12}\xi_{21}/\xi_{22}) 
% \nonumber 
% \\
&=& \om_q -\gam_1\frac{\cos(\om_ql_1)\sin[\om_q(l_2-l_1)]}{\cos(\om_ql_2)}.
\eea
This implies that,
owing to JQF,  
DQ acquires an infinite radiative lifetime
% JQF protects DQ from the radiative decay, 
regardless of its position $l_1(<l_2)$.

For comparison, we consider the case of $l_1 > l_2$. 
The renormalized complex frequency of DQ is then given by
\bea
\tom_1'' &=& \om_q - \gam_1 e^{i\om_q(l_1-l_2)}\sin[\om_q(l_1-l_2)],
\eea 
the imaginary part of which vanishes only when $\sin[\om_q(l_1-l_2)]=0$. 
Therefore, in contrast with the case of $l_1<l_2$,
the radiative decay of DQ is prohibited
only when the DQ locates exactly at 
$l_1=l_2 + n\lam_q/2$ ($n=0,1,\cdots$).

These results can be understood intuitively 
by assuming that JQF functions as an ^^ ^^ atomic mirror''
and imposes a closed boundary condition
on the waveguide field at $r=l_2$. 
Then, in the $0 \leq r \leq l_2$ region,
JQF forms an effective cavity 
whose eigenfrequencies are $\om^\mathrm{eff}_n=(\frac{1}{2}+n)\pi/l_2$. 
The decay of DQ is prohibited as long as $\om_q \neq \om^\mathrm{eff}_n$.
On the other hand, in the $r \geq l_2$ region, 
the eigenmode spectrum is continuous 
due to the semi-infiniteness of the waveguide. 
The vacuum fluctuation of the waveguide field 
at the qubit frequency $\om_q$
vanishes at $r=l_2 + n\lam_q/2$ ($n=0,1,\cdots$)
because the eigenmode function is a standing wave 
having a node at $r=l_2$.
The decay of DQ is prohibited only when it locates there.

The effect of the intrinsic decay of JQF 
other than the radiative decay into the line,
which is assumed to be absent throughout this paper, 
is discussed in Appendix~\ref{sec:gint}. 
It is revealed there that 
the radiative decay rate of DQ
is suppressed by a factor of $\gam_{\mathrm{i}2}/\gam_2$,
where $\gam_{\mathrm{i}2}$ ($\gam_2$) is 
the intrinsic (radiative) decay rate of JQF.
Therefore, the suppression becomes imperfect 
when JQF has a finite intrinsic decay rate, 
but is still substantial if $\gam_{\mathrm{i}2} \ll \gam_2$.

%%%%%%%%%%%%%%%%%%%%%%%%%%%%%%%%%%%%%%%%%%%%%%%%%%%%%%%%%%%%%%%%%%%%%%%%%%%%%%%
\subsection{Time evolution}%\label{sec:Ham}
%%%%%%%%%%%%%%%%%%%%%%%%%%%%%%%%%%%%%%%%%%%%%%%%%%%%%%%%%%%%%%%%%%%%%%%%%%%%%%%
%==============================================================================
\begin{figure}[t]
\begin{center}
\includegraphics[scale=1.0]{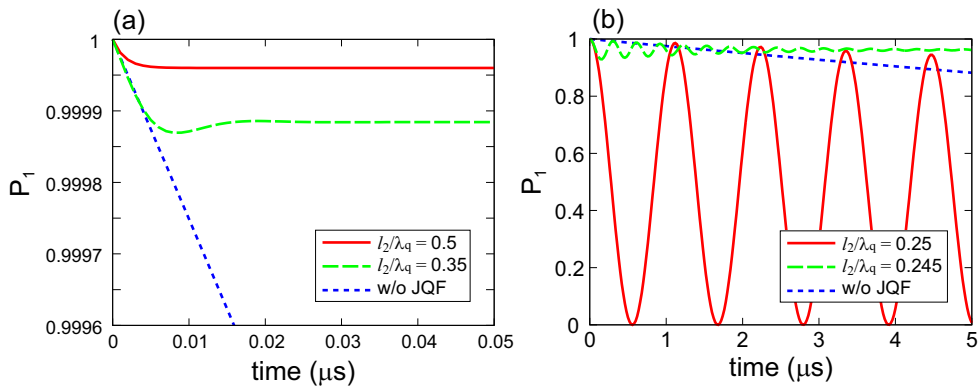}
\end{center}
\caption{Survival probability of DQ, $P_1(t)=|\alp_1(t)|^2$.
The parameter values are described in the caption of Fig.~\ref{fig:config}, 
except that the JQF position $l_2$ is varied.
(a)~Results for $l_2/\lam_q=0.5$ (red solid) and $0.35$ (green dashed).
(b)~Results for $l_2/\lam_q=0.25$ (red solid) and $0.245$ (green dashed).
The free survival probability in the absence of JQF
is also shown (blue dotted).
}
\label{fig:decay}
\end{figure}
%==============================================================================
The solution of Eqs.~(\ref{eq:da1dt}) and (\ref{eq:da2dt}) are given by
\bea
\alp_1(t) &=& 
\left( \frac{\mu_2+\xi_{11}}{\mu_2-\mu_1}e^{\mu_1 t} 
     + \frac{\mu_1+\xi_{11}}{\mu_1-\mu_2}e^{\mu_2 t}\right) e^{-i\om_q t},
\label{eq:a1}
\\
\alp_2(t) &=& 
\frac{\xi_{21}}{\mu_2-\mu_1}(e^{\mu_1 t}-e^{\mu_2 t})e^{-i\om_q t},
\label{eq:a2}
\eea
where $\mu_1$ and $\mu_2$ are defined by
\beq
(z+\xi_{11})(z+\xi_{22})-\xi_{12}\xi_{21}=(z-\mu_1)(z-\mu_2).
\label{eq:mu12}
\eeq
The survival probability of DQ is given by $P_1(t)=|\alp_1(t)|^2$.
In Fig.~\ref{fig:decay}, fixing the position of DQ at $l_1=0$, 
time evolution of $P_1(t)$ is shown for several values of $l_2$. 
Expectedly, we observe that the decay of DQ is suppressed,  
except when $l_2 \approx \frac{1+2n}{4}\lam_q$. 
The optimal position of JQF is $l_2=n\lam_q/2$. 
Equations~(\ref{eq:a1}) and (\ref{eq:a2}) then reduce to the following forms:
\bea
\alp_1(t) &=& 
\left(
\frac{\gam_2}{\gam_1+\gam_2}+ 
\frac{\gam_1}{\gam_1+\gam_2}e^{-(\gam_1+\gam_2)t}\right)e^{-i\om_q t},
\label{eq:a1v2}
\\
\alp_2(t) &=& (-)^{n+1}\frac{\sqrt{\gam_1\gam_2}}{\gam_1+\gam_2
}[1-e^{-(\gam_1+\gam_2)t}]e^{-i\om_q t}.
\label{eq:a2v2}
\eea
In the $t \to \infty$ limit, 
$P_1(t) = (\frac{\gam_2}{\gam_1+\gam_2})^2 
\approx 1-\frac{2\gam_1}{\gam_2}$. 
Therefore, the radiative decay of DQ is 
mostly prohibited when $\gam_2 \gg \gam_1$.

Such prohibited decay of DQ
originates in the subradiance effect~\cite{goban2015,ostermann2019,albrecht2019}. 
This is understood most clearly in the case of $l_1=l_2=0$. 
From the Hamiltonian of Eq.~(\ref{eq:Ham}), 
we observe that the following two states, 
$|\mathrm{sup}\ra$ and $|\mathrm{sub}\ra $, 
respectively correspond to the super- and sub-radiant states
within the one-excitation subspace of the two qubits: 
\bea
|\mathrm{sup}\ra 
&=& 
\frac{\sqrt{\gam_1}\hsig_1^{\dag}+\sqrt{\gam_2}\hsig_2^{\dag}}
{\sqrt{\gam_1+\gam_2}}
|v \ra,
\\
|\mathrm{sub}\ra 
&=& 
\frac{\sqrt{\gam_2}\hsig_1^{\dag}-\sqrt{\gam_1}\hsig_2^{\dag}}
{\sqrt{\gam_1+\gam_2}}
|v \ra.
\eea
The superradiant state decays rapidly with a rate of $2(\gam_1+\gam_2)$, whereas 
the subradiant state is an eigenstate of the Hamiltonian and does not decay.
The excited state of DQ is mostly composed of the subradiant state
and contains a tiny fraction of the superradiant state, 
\bea
\hsig_1^{\dag}|v \ra
&=& 
\frac{\sqrt{\gam_1}|\mathrm{sup}\ra
    + \sqrt{\gam_2}|\mathrm{sub}\ra}
{\sqrt{\gam_1+\gam_2}}.
\eea
The rapid decay of the superradiant component is observed as 
the initial drop of the survival probability in Fig.~\ref{fig:decay}(a).  
The quasi-stationary $P_1 \, [=\gam_2^2/(\gam_1+\gam_2)^2]$
results from the subradiant components.
These arguments are compatible with 
Eqs.~(\ref{eq:a1v2}) and (\ref{eq:a2v2}) with $n=0$. 

%%%%%%%%%%%%%%%%%%%%%%%%%%%%%%%%%%%%%%%%%%%%%%%%%%%%%%%%%%%%%%%%%%%%%%%%%%%%%%%
\section{Microwave response}\label{sec:mr}
%%%%%%%%%%%%%%%%%%%%%%%%%%%%%%%%%%%%%%%%%%%%%%%%%%%%%%%%%%%%%%%%%%%%%%%%%%%%%%%
In the previous section, 
we derived the radiative decay rate of DQ analytically
and confirmed that JQF protects DQ from radiative 
decay through the control line.
In this section, we numerically investigate the quantum control of DQ 
with a microwave pulse applied through the waveguide. 
We will observe that,
as long as the control pulse is sufficiently strong,
we can control DQ as if JQF is absent. 
This implies that JQF does not affect the gate time $T_g$ of DQ
while enhancing its lifetime $T_1$, % (Sec.~\ref{sec:rd}),
and thus breaks the trade-off relation of Eq.(\ref{eq:tradeoff}).

We assume that both qubits are in the ground state
at the initial moment ($t=0$), 
and a classical control field $E_\mathrm{in}(t)$ is applied 
through the waveguide for $t>0$.
The spatial waveform of the control field at $t=0$ is $E_\mathrm{in}(-r)$.
Therefore, the initial state vector is written as
\bea
|\phi(0)\ra &=& 
\cN \exp \left( \int_{-\infty}^0 \!\! dr \ E_\mathrm{in}(-r) \tb_r^{\dag} 
\right)|v \ra,
\label{eq:psi0}
\eea
where $\cN=\exp\left(-\int dr |E_\mathrm{in}(-r)|^2/2\right)$ 
is a normalization factor.
Note that this is in a coherent state and therefore is
an eigenvector of the noise operator, Eq.~(\ref{eq:Njt}).
Hereafter, we use the notation of $\la\phi(0)|\hat{A}(t)|\phi(0)\ra = \la\hat{A}(t)\ra$. 
From Eqs.~(\ref{eq:dsig1dt}) and (\ref{eq:psi0}), 
the equation of motion for $\la\hsig_1(t)\ra$ is given by
\bea
\frac{d}{dt}\la\hsig_1\ra &=& (-i\om_q-\xi_{11})\la\hsig_1\ra 
-\xi_{12}\la\hsig_2\ra
+2\xi_{12} \la\hsig_1^{\dag}\hsig_1\hsig_2\ra
+i(1-2\la\hsig_1^{\dag}\hsig_1\ra)\la\hN_1(t)\ra,
% \label{eq:dsig1dt}
\eea
where
\bea
\la \hN_j(t) \ra &=& 
\sqrt{2\gam_j}\cos(\om_ql_j) E_\mathrm{in}(t).
\label{eq:<Njt>}
\eea
We numerically solve the simultaneous differential equations 
of the following 9 quantities:
$\la \hsig_1 \ra$, 
$\la \hsig_2 \ra$, 
$\la \hsig_1^{\dag}\hsig_1 \ra$, 
$\la \hsig_2^{\dag}\hsig_2 \ra$, 
$\la \hsig_2^{\dag}\hsig_1 \ra$, 
$\la \hsig_1\hsig_2 \ra$, 
$\la \hsig_1^{\dag}\hsig_1\hsig_2 \ra$, 
$\la \hsig_2^{\dag}\hsig_1\hsig_2 \ra$, and 
$\la \hsig_1^{\dag}\hsig_2^{\dag}\hsig_1\hsig_2 \ra$.

%%%%%%%%%%%%%%%%%%%%%%%%%%%%%%%%%%%%%%%%%%%%%%%%%%%%%%%%%%%%%%%%%%%%%%%%%%%%%%%
\subsection{Rabi oscillation}\label{ssec:ro}
%%%%%%%%%%%%%%%%%%%%%%%%%%%%%%%%%%%%%%%%%%%%%%%%%%%%%%%%%%%%%%%%%%%%%%%%%%%%%%%
%==============================================================================
\begin{figure}[t]
\begin{center}
\includegraphics[scale=1.0]{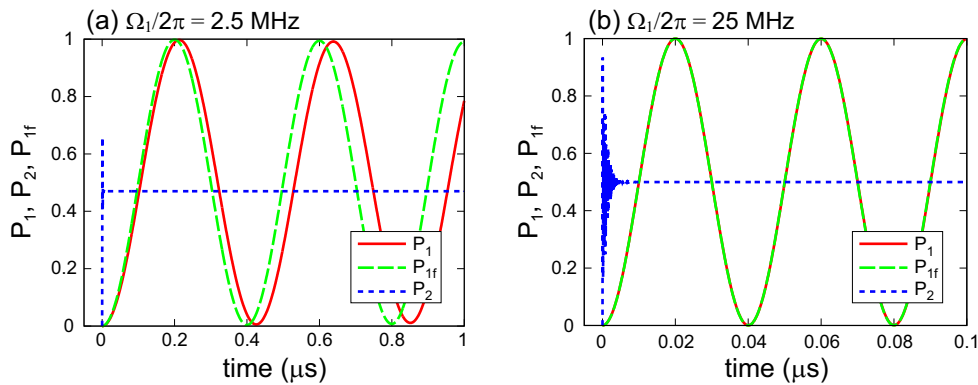}
\end{center}
\caption{
Rabi oscillations of DQ ($P_1$, red solid) and JQF ($P_2$, blue dotted).
The free Rabi oscillation of DQ ($P_{1f}$, green dashed) in the absence of JQF
is also shown for reference.
The Rabi frequency is set at $\Om_1/2\pi=2.5$~MHz in (a), and $25$~MHz in (b).
The corresponding photon rates are 
$|E_d|^2/\gam_2=3.91$ and $391$, respectively.
$P_1$ and $P_{1f}$ are almost overlapping in (b).
}
\label{fig:Rabi}
\end{figure}
%==============================================================================
First, we investigate the case of continuous drive,
\bea
E_\mathrm{in}(t) &=&
\begin{cases}
0 & (t<0) \\
E_d e^{-i\om_d t} & (t \geq 0)
\end{cases},
\eea
and observe the Rabi oscillations of DQ and JQF induced by this drive field.
Note that $|E_d|^2$ represents the incoming photon rate. 
In the numerical simulations, 
we assume that both DQ and JQF are placed at their optimal positions
($l_1=0$, $l_2=\lam_q/2$) and that 
a resonant drive field ($\om_d = \om_q$) is applied through the waveguide.

In Fig.~\ref{fig:Rabi}, the excitation probabilities of DQ and JQF,
$P_j(t)=\la \hsig_j^{\dag}\hsig_j \ra$,  
are plotted for various drive power. 
In order to emphasize the effect of JQF on DQ, 
we also plot the {\it free} Rabi oscillation of DQ, $P_{1f}(t)$, 
assuming the absence of JQF. 
$P_{1f}(t)$ is analytically given by
\bea
P_{1f}(t) &=& 
\frac{\Om_1^2}{2(\Om_1^2+2\gam_1^2)}
\left[ 1-e^{-3\gam_1 t/2}\left(
\cos\tOm_1 t + \frac{3\gam_1}{2\tOm_1}\sin\tOm_1 t
\right) \right],
\label{eq:s1s1}
\eea
where $\Om_1=\sqrt{8\gam_1|E_d|^2}$ (Rabi frequency) 
and $\tOm_1=\sqrt{8\gam_1|E_d|^2-(\gam_1/2)^2}$. 
We observe that $P_{1}(t)$ and $P_{1f}(t)$ are almost overlapping
for a strong drive satisfying $\gam_2 \ll |E_d|^2$. 
Thus, the Rabi oscillation of DQ is almost unaffected by JQF. 
This is explained as follows: 
In the right-hand-side of Eq.~(\ref{eq:dsig1dt}),
we see that DQ is driven by two different mechanisms:
the mutual interaction between DQ and JQF (second term) and 
the applied field (third term). 
Therefore, the condition that 
the Rabi oscillation of DQ is unaffected by JQF
is $|\xi_{12}| \ll \la\hN_1(t)\ra$, 
which reduces to $\gam_2 \ll |E_d|^2$.

On the other hand, JQF undergoes a much faster Rabi oscillation than DQ
($\Om_2/\Om_1 = \sqrt{\gam_2/\gam_1} \approx 224$),
and relaxes to its stationary value of $\Om_2^2/2(\Om_2^2+2\gam_2^2)$
within a relaxation time of $\gam_2^{-1} \sim 5$~ns.
For a strong field satisfying $|E_d|^2 \gg \gam_2$,
JQF becomes saturated, and the stationary state is 
approximately the maximally mixed state.

%%%%%%%%%%%%%%%%%%%%%%%%%%%%%%%%%%%%%%%%%%%%%%%%%%%%%%%%%%%%%%%%%%%%%%%%%%%%%%%
\subsection{$\pi$-pulse excitation}%\label{sec:Ham}
%%%%%%%%%%%%%%%%%%%%%%%%%%%%%%%%%%%%%%%%%%%%%%%%%%%%%%%%%%%%%%%%%%%%%%%%%%%%%%%
%==============================================================================
\begin{figure}
\begin{center}
\includegraphics[scale=1.0]{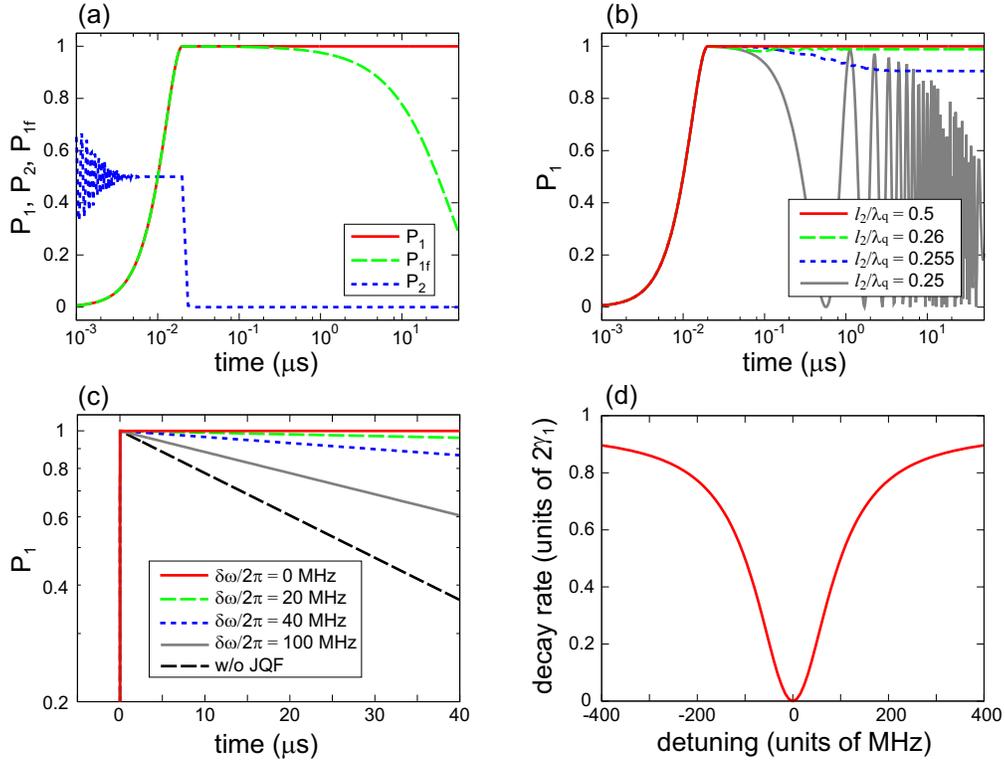}
\end{center}
\caption{
Response to $\pi$-pulse. The pulse is on for $0 \leq t \leq 20$~ns.
(a)~Excitation probability of DQ ($P_1$, red solid) and JQF ($P_2$, blue dotted).
The former in the absence of JQF ($P_{1f}$, green dashed) is also plotted. 
$P_{1}$ and $P_{1f}$ are overlapping for $t \lesssim 0.2\mu\mathrm{s}$.
(b)~$P_1(t)$ for various JQF position $l_2$. 
(c)~$P_1(t)$ for various JQF frequency $\om_2$.
The detuning $\delta\om$ in the legend is defined by $\delta\om=\om_2-\om_1$.
(d)~Dependence of the radiative decay rate of DQ on the detuning.
}
\label{fig:g2dep}
\end{figure}
\begin{figure}% [b]
\begin{center}
\includegraphics[scale=1.0]{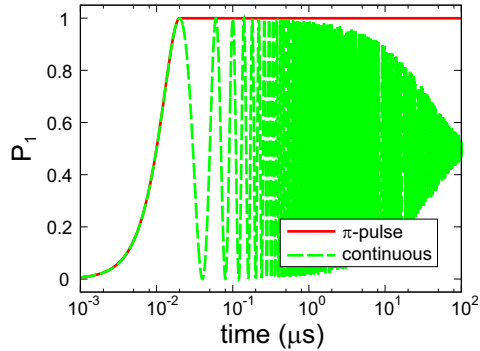}
\end{center}
\caption{
Comparison of continuous-wave and $\pi$-pulse excitations.
The Rabi frequency is set at $\Om_1/2\pi=25$~MHz and 
the duration of $\pi$ pulse is set at $20$~ns.
}
\label{fig:cp}
\end{figure}

%==============================================================================
Here we observe the dynamics of DQ and JQF induced by a $\pi$-pulse for DQ. 
We employ a square $\pi$-pulse with a length of 20~ns ($\Om_1/2\pi=25$~MHz). 
Time evolution of the excitation probabilities of DQ ($P_1$) 
and JQF ($P_2$) are plotted in Fig.~\ref{fig:g2dep}(a). 
JQF exhibits a rapid damped Rabi oscillation 
and relaxes to the maximally mixed state ($P_2=1/2$)
during $\pi$-pulse irradiation. 
After the $\pi$-pulse is switched off, 
JQF quickly decays to the ground state.
On the other hand,
DQ is excited by the $\pi$-pulse and keeps staying in the excited state 
even after the $\pi$-pulse is switched off.
The stationary value of $P_1$ is 0.9997. 
As we observe later (Fig.~\ref{fig:cp}), 
damping of DQ is not suppressed completely while the control field is on. 
The tiny unexcited probability 0.00032 is mainly due to 
such damping during $\pi$-pulse irradiation.

The JQF position $l_2$ is varied in Fig.~\ref{fig:g2dep}(b).
A remarkable fact here is that
the decay of DQ is prohibited
even when JQF is not placed exactly at its optimal position:
the stationary excitation probability reaches 0.9994 for $l_2/\lam_q=0.35$ (non-optimal), 
which is comparable to 0.9997 for $l_2/\lam_q=0.5$ (optimal). 
Therefore, the precise positioning of JQF is unnecessary for protection of DQ. 
This makes practical implementation of the present scheme easier.

The JQF frequency $\om_2$ is varied in Fig.~\ref{fig:g2dep}(c).
We observe that, in the presence of detuning, 
the protection of DQ by JQF is imperfect, 
resulting in the exponential decay of DQ.
In Fig.~\ref{fig:g2dep}(d), 
the DQ decay rate is plotted as a function of the detuning. 
The functional form of this decay rate is given by Eq.~(\ref{eq:gamDQ}).
The width of the dip in the decay rate is about 200~MHz, 
which is determined by the JQF linewidth $\gam_2$. 

%==============================================================================
%==============================================================================
Continuous-wave and $\pi$-pulse excitations are compared in Fig.~\ref{fig:cp}.
As we observed in Eq.~(\ref{eq:s1s1}), % Sec.~\ref{ssec:ro}, 
under continuous-wave excitation, 
DQ exhibits damped Rabi oscillation with a decay rate of $3\gam_1/2$.
This implies that, while the control field is on, 
the control line works as a dissipation channel and damps the dynamics of DQ.  
In contrast, after the $\pi$-pulse excitation, 
DQ stays in the excited state without being dissipated.

\begin{figure}%[b]
\begin{center}
\includegraphics[scale=0.95]{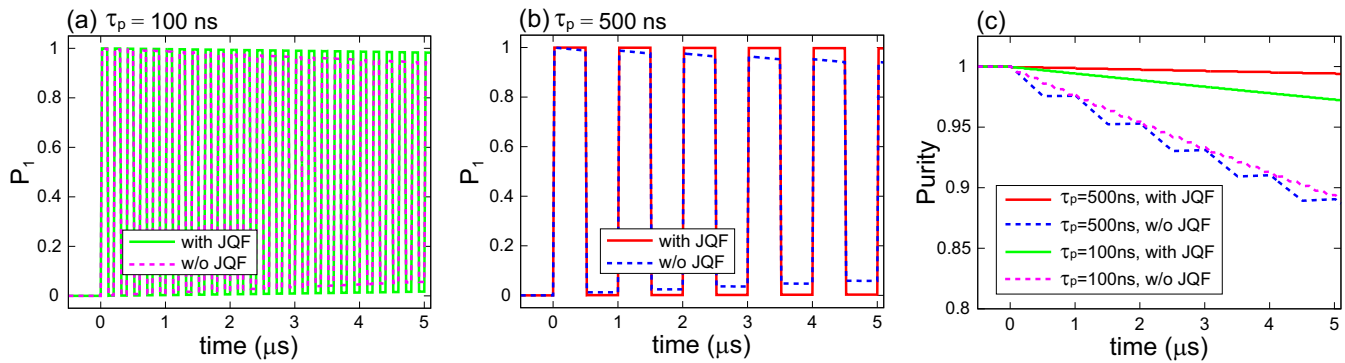}
\end{center}
\caption{
Successive application of $\pi$-pulses. 
The $\pi$-pulse duration time is 20~ns.
(a)~Time evolution of the excitation probability of DQ
with the pulse period $\tau_p=100$~ns.
The green solid (magenta dotted) line shows the results with (without) JQF.
(b)~The same plot as (a) for $\tau_p=500$~ns.
The red solid (blue dotted) line shows the results with (without) JQF.
(c)~Time evolution of the purity of DQ.
The upper (lower) two lines show the results with (without) JQF. 
}
\label{fig:suc}
\end{figure}

In Fig.~\ref{fig:suc}, we show the bit flips of DQ 
between the ground and excited states induced by successive $\pi$-pulses. 
The $\pi$-pulse duration time is set at 20~ns,  
and the pulse period $\tau_p$ is set at 100~ns (duty ratio 0.2) 
in Fig.~\ref{fig:suc}(a) and at 500~ns (duty ratio 0.04) in Fig.~\ref{fig:suc}(b). 
Without JQF, the amplitude of $P_1$ oscillation 
damps with a rate around $\gam_1$: 
$P_1$ at $t=5.02~\mu$s is 0.9436 in (a) and 0.9417 in (b).
The damping rate is almost insensitive to the duty ratio,
since damping of DQ always occurs, 
regardless of the control pulse is on or off.
In the presence of JQF, damping of DQ is substantially suppressed: 
$P_1$ at $t=5.02~\mu$s reaches 0.9828 in (a) and 0.9968 in (b).
Operation with lower duty ratio is more advantageous, 
since JQF protects DQ only during the control pulse is off.   
This is also confirmed in Fig.~\ref{fig:suc}(c),
which plots time evolution of the purity of DQ, 
${\cal P}=\la\hsig_1^{\dag}\hsig_1\ra^2 + 
(1-\la\hsig_1^{\dag}\hsig_1\ra)^2+2|\la\hsig_1\ra|^2$.

%==============================================================================
%==============================================================================

%%%%%%%%%%%%%%%%%%%%%%%%%%%%%%%%%%%%%%%%%%%%%%%%%%%%%%%%%%%%%%%%%%%%%%%%%%%%%%%
\section{summary}\label{sec:sum}
%%%%%%%%%%%%%%%%%%%%%%%%%%%%%%%%%%%%%%%%%%%%%%%%%%%%%%%%%%%%%%%%%%%%%%%%%%%%%%%
For scalable quantum information processing, 
we should perform fast gate operations on the qubits,
while keeping long coherence times of the qubits. 
However, there exists a trade-off between them, 
which cannot be resolved by a conventional Purcell filter.
In this work, we proposed a Josephson quantum filter (JQF), 
which protects a data qubit (DQ) 
from radiative decay into the control line without losing the gate speed. 
In the proposed setup, 
DQ is coupled to an end of a semi-infinite control line 
and JQF is coupled to the same line 
with a distance from DQ of the order of resonance wavelength. 
Owing to a superradiance effect, 
JQF suppresses radiative decay of DQ under the following conditions: 
(i)~DQ and JQF are resonant, 
(ii)~JQF couples to the control line by far more strongly than DQ, 
and (iii)~the DQ--JQF distance is close to 
integer multiples of half of the resonance wavelength. 
We numerically confirmed that 
the speed of the gate operations on DQ is unaffected by JQF. 
The radiative decay of DQ is completely suppressed 
by JQF when the control field is off, 
whereas it is not under irradiation of the control field.
Thus, the operation of JQF is somewhat similar to that of a tunable coupler
between a qubit and a control line.
However, JQF is a passive element that is free from active control on it,
and therefore is highly suitable for integration %ed to be incorporated 
in complicated circuits with many qubits.

%%%%%%%%%%%%%%%%%%%%%%%%%%%%%%%%%%%%%%%%%%%%%%%%%%%%%%%%%%%%%%%%%%%%%%%%%%%%%%%
\section*{Acknowledgments}% \label{sec:sum}
%%%%%%%%%%%%%%%%%%%%%%%%%%%%%%%%%%%%%%%%%%%%%%%%%%%%%%%%%%%%%%%%%%%%%%%%%%%%%%%
We acknowledge the fruitful discussions with S. Masuda, T. Shitara,
and J. Gea-Banacloche. 
This work was supported in part by 
JSPS KAKENHI (no. 19K03684 and no. 26220601), 
JST ERATO (grant no. JPMJER1601), UTokyo ALPS, and MEXT Q-LEAP.

\appendix
%%%%%%%%%%%%%%%%%%%%%%%%%%%%%%%%%%%%%%%%%%%%%%%%%%%%%%%%%%%%%%%%%%%%%%%%%%%%%%%
\section{Validation of free-evolution approximation}\label{sec:app}
%%%%%%%%%%%%%%%%%%%%%%%%%%%%%%%%%%%%%%%%%%%%%%%%%%%%%%%%%%%%%%%%%%%%%%%%%%%%%%%
In this section, 
in order to check the validity of the free-evolution approximation, 
we analyze the radiative decay of DQ 
without the free-evolution approximation. 
The initial state vector [Eq.~(\ref{eq:inivec})] and 
the state vector at time $t$ [Eq.~(\ref{eq:tvec})]
remain unchanged in the rigorous analysis.
However, without the free-evolution approximation, 
the equations of motion for $\alp_1(t)$ and $\alp_2(t)$ become 
delay differential equations.
Instead of Eqs.~(\ref{eq:da1dt}) and (\ref{eq:da2dt}), 
they are given by
\bea
\frac{d}{dt}\alp_1(t) &=& 
\left(-i\om_q-\frac{\gam_1}{2}\right)\alp_1(t)-\frac{\gam_1}{2}\alp_1(t-2l_1)
-\frac{\sqrt{\gam_1\gam_2}}{2}[\alp_2(t-l_1-l_2)+\alp_2(t+l_1-l_2)],
\label{eq:da1dt_v2}
\\
\frac{d}{dt}\alp_2(t) &=& 
\left(-i\om_q-\frac{\gam_2}{2}\right)\alp_2(t)-\frac{\gam_2}{2}\alp_2(t-2l_2)
-\frac{\sqrt{\gam_1\gam_2}}{2}[\alp_1(t-l_1-l_2)+\alp_1(t+l_1-l_2)],
\label{eq:da2dt_v2}
\eea
with the initial conditions of $\alp_1(0)=1$ and $\alp_2(0)=0$. 
Note that $\alp_{1,2}(t)=0$ for $t<0$.

\begin{figure}%[b]
\begin{center}
\includegraphics[scale=1.0]{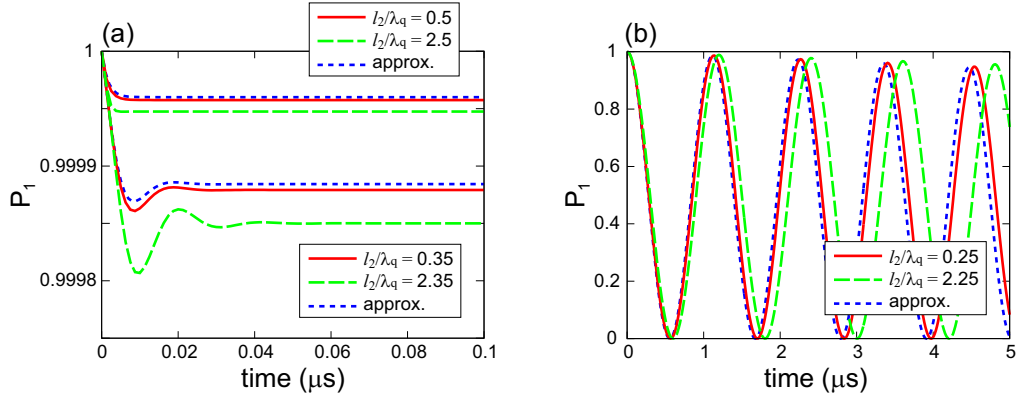}
\end{center}
\caption{
Comparison of the rigorous survival probability 
and the approximate ones based on the free-evolution approximation. 
}
\label{fig:check}
\end{figure}

In Fig.~\ref{fig:check}, fixing $l_1(=0)$ and varying $l_2$, 
we compare the rigorous and approximate survival probabilities $P_1(t)=|\alp_1(t)|^2$.
The upper three lines in Fig.~\ref{fig:check}(a) represent
the rigorous ones for $l_2/\lam_q=0.5$ and $2.5$ 
and the approximate one. 
Note that the free-evolution approximation yields
the same $P_1(t)$ for $l_2/\lam_q=0.5$ and $2.5$.  
As expected, the agreement between the rigorous and approximate results 
becomes better for a shorter round-trip time $2l_2$. 
The deviation between the rigorous and approximate $P_1(t)$ 
is negligibly small for $l_2/\lam_q=0.5$
($2.6 \times 10^{-6}$ in the stationary state), 
indicating the validity of the approximation. 
The lower three lines in Fig.~\ref{fig:check}(a) and Fig.~\ref{fig:check}(b)
show the results for the non-optimal filter position. 
The deviation between the rigorous and approximate results are larger 
in comparison with the optimal cases, 
but the approximation is fairly good for $l_2 \lesssim \lam_q$.

%==============================================================================
%==============================================================================

%%%%%%%%%%%%%%%%%%%%%%%%%%%%%%%%%%%%%%%%%%%%%%%%%%%%%%%%%%%%%%%%%%%%%%%%%%%%%%%
\section{Effect of intrinsic loss and detuning}\label{sec:gint}
%%%%%%%%%%%%%%%%%%%%%%%%%%%%%%%%%%%%%%%%%%%%%%%%%%%%%%%%%%%%%%%%%%%%%%%%%%%%%%%
Here, we derive the decay rate of DQ 
in the presence of the intrinsic loss of DQ and JQF
and the detuning between them.
We denote the intrinsic decay rates of DQ and JQF by
$\gam_\mathrm{i1}$ and $\gam_\mathrm{i2}$, respectively.
Then, Eqs.~(\ref{eq:da1dt}) and (\ref{eq:da2dt}) are modified as follows:
\bea
\frac{d\alp_1}{dt} 
&=& 
-(i\om_1+\xi_{11}+\gam_\mathrm{i1}/2)\alp_1 -\xi_{12}\alp_2,
\label{eq:da1dt2}
\\
\frac{d\alp_2}{dt} 
&=& 
-\xi_{21}\alp_1 -(i\om_2+\xi_{22}+\gam_\mathrm{i2}/2)\alp_2. 
\label{eq:da2dt2}
\eea
Switching to the frame rotating at $\om_1$ 
and solving Eq.~(\ref{eq:da2dt2}) adiabatically, 
the complex frequency of DQ is given by
\bea
\tom_1' 
&=& \om_q + i\left(
\frac{\xi_{12}\xi_{21}}{\xi_{22}+\gam_\mathrm{i2}/2+i\delta\om}
-\xi_{11}-\frac{\gam_\mathrm{i1}}{2}
\right),
\eea
where $\delta\om=\om_2-\om_1$ is the detuning between DQ and JQF.
The decay rate of DQ is determined by 
$\gam_{DQ}=-2\mathrm{Im}(\tom_1)$.
For the optimal case of $l_1=0$ and $l_2=n\lam_q/2$
($n=0,1,\cdots$), $\gam_{DQ}$ reduces to the following form: 
\bea
\gam_{DQ}
&=&
\gam_\mathrm{i1}+2\gam_1
\left[
1-\frac{\gam_2(\gam_2+\gam_\mathrm{i2}/2)}{(\gam_2+\gam_\mathrm{i2}/2)^2+(\delta\om)^2}
\right].
\label{eq:gamDQ}
\eea
The first term represents the intrinsic decay rate of DQ, 
which is unaffected by JQF as expected. 
The second term represents the radiative decay rate of DQ, 
which is suppressed by JQF. 
In the absence of detuning and in the $\gam_2 \gg \gam_\mathrm{i2}$ limit, 
the radiative decay rate of DQ is approximately given by
$\gam_1 \times (\gam_\mathrm{i2}/\gam_\mathrm{2})$. 
This implies that,
even when JQF has intrinsic loss, 
we can substantially suppress the radiative decay of DQ
by making the radiative decay of JQF dominant.
Equation~(\ref{eq:gamDQ}) agrees with the Lorentzian dip 
observed in Fig.~\ref{fig:g2dep}(d).

%%%%%%%%%%%%%%%%%%%%%%%%%%%%%%%%%%%%%%%%%%%%%%%%%%%%%%%%%%%%%%%%%%%%%%%%%%%%%
%%%%%%%%%%%%%%%%%%%%%%%%%%%%%%%%%%%%%%%%%%%%%%%%%%%%%%%%%%%%%%%%%%%%%%%%%%%%%

%%%%%%%%%%%%%%%%%%%%%%%%%%%%%%%%%%%%%%%%%%%%%%%%%%%%%%%%%%%%%%%%%%%%%%%%%%%%%
%%%%%%%%%%%%%%%%%%%%%%%%%%%%%%%%%%%%%%%%%%%%%%%%%%%%%%%%%%%%%%%%%%%%%%%%%%%%%%
%%%%%%%%%%%%%%%%%%%%%%%%%%%%%%%%%%%%%%%%%%%%%%%%%%%%%%%%%%%%%%%%%%%%%%%%%%%%%%
\end{document}